\documentstyle[eqsecnum,preprint,aps]{revtex}
\begin{document}
\title{Growth Kinetics for a System with a \\  Conserved Order Parameter: \\
Off Critical Quenches}
\author{Gene F. Mazenko and Robert A. Wickham}
\address{The James Franck Institute and the Department of Physics\\ The
University of Chicago\\ Chicago, Illinois 60637}
\date{cond-mat/9411045}
\maketitle
\begin{abstract}
The theory of growth kinetics developed previously is extended to the
asymmetric case of off-critical quenches for systems with a conserved scalar
order parameter. In this instance the new parameter $M$, the average global
value of the order parameter, enters the theory. For $M=0$ one has critical
quenches, while for sufficiently large $M$
one approaches the coexistence curve. For all $M$ the theory supports a scaling
solution for the order parameter correlation function with the
Lifshitz-Slyozov-Wagner growth law $L \sim t^{1/3}$. The theoretically
determined scaling function depends only on the spatial dimensionality $d$ and
the parameter $M$, and is determined explicitly here in two and three
dimensions. Near the coexistence curve oscillations in the scaling function are
suppressed. The structure factor displays Porod's law $Q^{-(d+1)}$ behaviour at
large scaled wavenumbers $Q$, and $Q^{4}$ behaviour at small scaled
wavenumbers, for all $M$. The peak in the structure factor  widens as $M$
increases and  develops a significant tail for quenches near the coexistence
curve.  This is in  qualitative agreement with simulations.
\end{abstract}

\section{Introduction}

In previous work \cite{maz} a theory based on a field-theoretic Langevin model
was developed to treat the growth kinetics of a system with a conserved scalar
order parameter for the case of symmetric or critical quenches. In this paper
the lowest order version of this  theory is extended to off-critical quenches.
Quenches to a final state near the coexistence curve where the volume fraction
of the minority phase is small have been studied by a variety of approaches,
but theoretical studies of the Langevin model have never been extended into
this regime. The techniques developed thus far are generalizations of the
Lifshitz-Slyosov-Wagner (LSW) treatment \cite{LSW,Wagner} which considers one
spherical droplet interacting with a mean concentration field. This approach is
valid only in the limit of zero volume fraction,
but other mean field theoretic and statistical mechanical techniques have been
developed to incorporate the effects of the interaction of other droplets and
extend the theory to slightly larger volume fractions
\cite{Rikvold,Marqusee,Tomita-model,Ohta,Tokuyama,Furukawa,Marder,Yao,Nakahara}
{}. Another approach to the problem is to use numerical simulations in concert
 with a theory describing the concentration field around spherical droplets
(essentially an electrostatics problem with moving boundary conditions)
\cite{Voorhees}. Direct simulations of the Langevin equation exist for the
off-critical case in two dimensions \cite{Desai,chak}, but we are not aware of
 any such simulations for three dimensions \cite{KIsing}.

The theory developed in \cite{maz} shows how one can solve some of the thorny
problems associated with growth kinetics for the conserved order parameter
(COP) case. The theory can be evaluated as a well-defined sequence of
approximations with qualitative and quantitative improvement as one moves along
this sequence \cite{PGA1}. In this paper we limit ourselves to the lowest order
approximation in this formalism. From the work in \cite{maz} we know that there
are substantial limitations associated with this approximation and these are
discussed in some detail in section VI. However, it is also known from
\cite{maz} that this approximation gives good results for the scaling function
for correlations of the order parameter. We therefore concentrate on this
quantity here.

The new element in this work compared to \cite{maz} is that the average value
of the scalar order parameter $\psi$ is no longer zero:
\begin{equation}
\langle \psi ( {\bf R} , t) \rangle = M.
\end{equation}
$M$ is independent of ${\bf R}$ and $t$ because of the statistical homogeneity
of the system and the conservation law, respectively.

The main results of this paper are that, as in the critical case, the theory
supports a long-time scaling solution for the order parameter correlation
function
\begin{eqnarray}
C ( { \bf R},t) & = & \langle \delta \psi ( { \bf R}, t) \delta \psi ( { \bf 0
} , t) \rangle \nonumber \\
& = & \psi_{0}^{2} F \left( \frac{| { \bf R} | }{L(t)} \right)
\end{eqnarray}
where $ \delta \psi = \psi - M$, $\psi_{0}$ is the magnitude of the ordered
field in equilibrium, and $ L(t) $ is the characteristic length in the theory,
the average size of the domains. For later convenience we define the normalized
quantity $\tilde{M} = M/ \psi_{0}$, which ranges between $-1 \mbox{ and } 1$.
When $\tilde{M} = \pm 1$ the system is at the coexistence curve.  It is found
that, for all $\tilde{M}$ and for long times $t$ after the quench,  the growth
law $ L \sim t^{1/3} $ holds.  For small scaled distances $x$ ($= |{ \bf
R}|/L(t) $) one is able to find a scaling solution of the form
\begin{equation}
F(x) = 1 - \tilde{M}^{2} - e^{-y^{2}/2} \alpha x(1 + \beta_{2} x + \cdots)
\label{eq:smx}
\end{equation}
where the parameter $y$ is related to $\tilde{M}$ by
\begin{equation}
\tilde{M} = erf(y/\sqrt{2}).
\end{equation}
Unlike the non-conserved order parameter (NCOP) case treated earlier
\cite{TUG}, the coefficient $ \beta_{2} $ is not zero and must be determined,
along with $ \alpha$, as part of a non-linear eigenvalue problem. $ \beta_{2} $
is found to be negative for $ \tilde{M} = 0 $ and monotonically decreases as $
\tilde{M} \rightarrow 1 $. Thus, in this theory, the Tomita sum rule
($\beta_{2} = 0$) \cite {Tomita} is strongly broken as one approaches the
coexistence curve. This appears to be an important limitation of the current
theory.

The scaling function, $F$, has been determined explicitly in two and three
dimensions by solving the non-linear eigenvalue problem mentioned above. The
dependence of $F$ on $ \tilde{M} $ is weak for small $ \tilde{M} $. As $
\tilde{M} $ increases the first zero of $F$ moves to larger scaled distances
and the first minimum of $F$ becomes shallower. As $ \tilde{M} $ increases
further the oscillatory behaviour is suppressed and the predominant behaviour
is that of decay, as predicted by our large $y$ asymptotic analysis.  For large
$ \tilde{M} $ there are oscillations at large $x$ whose wavelength increases as
one approaches the coexistence curve. These oscillations preserve the
conservation law
\begin{equation}
\int d^{d} x F(x) = 0 \label{eq:conslaw}
\end{equation}
despite the existence of the strong decay.

The structure factor is the Fourier transform of the correlation function and
one has
\begin{equation}
C({\bf q},t) = \psi_{0}^{2} L^{d}(t) \tilde{F}(Q)
\end{equation}
where $Q = qL$ is a scaled wavenumber. $\tilde{F}(Q)$ is characterized by five
parameters. Since $\tilde{F}(Q)$ is a peaked quantity, the peak position
$Q_{max}$, and height $\tilde{F}(Q_{max})$, are of interest as functions of
$\tilde{M}$. The full-width at half-maximum, measured in units of $Q_{max}$, is
also relevant. The linear term in (\ref{eq:smx}) leads to Porod's law
\cite{Porod} for the large $Q$ tail of the structure factor
\begin{equation}
\tilde{F} (Q) = \tilde{F}(Q_{max}) A_{P}(\tilde{M}) \left(\frac{Q}{Q_{max}}
\right)^{-(d+1)},
\end{equation}
while for small $Q$, $\tilde{F}(Q)$ behaves as \cite{Q4}
\begin{equation}
\tilde{F} (Q) = \tilde{F}(Q_{max})A_{4}(\tilde{M}) \left( \frac{Q}{Q_{max}}
\right)^{4}.
\end{equation}
This  behaviour at small $Q$ is a result of  a conserved diffusive field
existing away from the interfaces which mediates the interaction among the
interfaces.
Our analysis shows that both $Q_{max}$ and $\tilde{F}(Q_{max})$ decrease to
zero as $\tilde{M} \rightarrow 1$. The width of the peak increases slightly for
small $ \tilde{M} $, but then develops a significant tail as $ \tilde{M}
\rightarrow 1$ for intermediate values of $Q$ near the base of the peak.
$A_{P}(\tilde{M})$ is a decreasing function of $\tilde{M}$, approaching zero in
a cusp as $\tilde{M} \rightarrow 1$. The coefficient $A_{4}(\tilde{M})$
increases with increasing $\tilde{M}$, growing rapidly near $\tilde{M} = 1$.
Damped oscillations are also seen in the structure factor around the $Q^{-4} $
behaviour at intermediate values of $Q$, a result also seen by other
investigators \cite{polymer study,Voorhees}.

When we compare our results with those of other workers we find good
qualitative agreement for both $F(x)$ and $\tilde{F}(Q)$ as  functions of
$\tilde{M}$. There are some quantitative differences, though. We believe that
the difference in the form of $\tilde{F}(Q)$ is due to our low estimate for the
coefficient $A_{4}$, leading to a peak in $\tilde{F}(Q)$ which is too narrow.
It seems likely that the lack of quantitative agreement is associated with the
breaking of the Tomita sum rule in the theory. On the other hand, this is the
only theory which has led to a determination of $A_{4}$, and one can hope that
using higher order approximations will give improved results.

In the next section the theory forming the basis for this work is outlined. In
section III the averages are performed which are relevant to the off-critical
case. The end result of these manipulations is a non-linear equation for the
scaling function. Section IV looks at the various limiting cases of the theory:
the small $x$, large $x$, small $Q$, and large $y$ behaviour. Section V
presents the numerical study of the equation for $F(x)$ in two and three
dimensions. Comparison of the results of this paper with results from other
investigators is made in section VI. The paper concludes with some comments
about future areas of research and improvements to the theory.

\newcommand{\be}{\begin{equation}}
\newcommand{\ee}{\end{equation}}

\section{The Model}

The dynamics are modelled using a noiseless time-dependent Ginzburg-Landau
equation for a conserved scalar order parameter $ \psi$ with a non-zero average
$M$:
\be
\frac{ \partial \psi(1)}{ \partial t_{1}} = D_{0} \nabla^{2}_{1} [ V'( \psi(1))
- \nabla^{2}_{1} \psi(1) ].   \label{eq:tdgl}
\ee
$V( \psi) $ is a double-well potential with degenerate minima at $ \pm
\psi_{0} $, but is otherwise unspecified since, as we shall see, our results
are independent of the precise form of $V$ \cite{asym pot}. Here, the
convienient notation in which 1 represents  $({\bf R_{1}},t_{1})$ is used. The
$D_{0} \nabla^{2}_{1}$ factor in (\ref{eq:tdgl}) ensures that the system has a
conserved order parameter. Thermal sources of noise are neglected because it is
assumed here that the quench is to zero temperature. Randomess enters the
problem through the initial conditions where we assume that the initial values
of  $\delta \psi = \psi - M $ are  governed by a Gaussian probability
distribution characterized by
\begin{eqnarray}
\langle \psi ({ \bf R},0) \rangle & = & M \\
\langle \delta \psi ({ \bf R},0) \delta \psi ({ \bf R'},0) \rangle & = &
\epsilon_{0} \delta ( { \bf R - R'}).
\end{eqnarray}
Our final results are independent of the amplitude $\epsilon_{0}$ appearing in
the initial distribution.

A method for extracting the correlation functions from (\ref{eq:tdgl}) is
described in \cite{maz}. Here we will merely outline the salient points. The
order parameter
is written as
\be
\psi(1) = \sigma[m(1)] - u(1) \label{eq:decomposition}
\ee
where $ \sigma$ is the equilibrium interfacial profile and $u$ represents
fluctuations about this ordered value. $m$ is assumed to be a random field
whose zeros correspond to the zeros of $ \sigma $, that is, to the positions of
the interfaces. The nature and interpretation  of $m$ will be discussed below.
In the NCOP case the fluctuating field $u$ can be safely ignored but in the COP
case it is the field $u$ which couples distant interfaces by permitting
currents of minority phase atoms to flow through the matrix. In \cite{maz} the
theory was closed by relating  $u$ back to $ \sigma$ and $m$ via the equation
\be
u(1) = \frac{u_{0} }{L} \sigma(1) + \lambda \nabla^{2}_{1} m(1) + \cdots
\label{eq:const}
\ee
where $u_{0} \mbox{ and } \lambda$ are parameters. This form satisfies the
desired properties that $u$ is conserved in
bulk, odd in $m$, and ${ \cal{O} } ( 1/L)$ everywhere. This last requirement
comes from the fact that the interfaces are a source of $u$ with a contribution
proportional to the local curvature of an interface $ \kappa \mbox{ } ( u|_{S}
\sim \kappa).$ It can then be shown that if (\ref{eq:const}) holds, $u$
satisfies the familiar form $ \nabla^{2} u = 0 $ away from the interfaces. $
\sigma$ is chosen to satisfy the equation for an equilibrium interface
\be
\frac{1}{2} \sigma_{2}(1) = V'( \sigma(m(1)) ) \label{eq:eqint}
\ee
where the factor 1/2 is inserted for convenience and $ \sigma_{n}(m) =
\partial^{n} \sigma(m) / \partial m^{n} $. The boundary condition, $\lim_{m
\rightarrow \pm \infty} \sigma = \pm \psi_{0}$, guarantees that the system
orders at the appropriate equilibrium value of the order parameter and results
in the useful relation
\be
\int^{\infty}_{- \infty} dx \sigma_{1} (x) = 2 \psi_{0} \label{eq:intsig1}.
\ee
As shown in \cite{maz}, equations (\ref{eq:const}) and (\ref{eq:eqint}) can be
substituted into (\ref{eq:tdgl}) and the result  multiplied by $ \sigma(2)$ and
averaged to get an equation for
\be
C_{ \sigma \sigma} ({ \bf R},t) = \langle \sigma(1) \sigma(2) \rangle.
\ee
At late times $1/L$ is expected to be small and one finds to leading order in
$1/L$:
\be
\frac{1}{2} \frac{ \partial}{ \partial t} C_{ \sigma \sigma}({ \bf R},t) = -
D_{0} q_{0}^{2} \nabla^{2} \left( \frac{u_{0}}{L} C_{ \sigma \sigma}({ \bf
R},t) + \lambda \nabla^{2} C_{ \sigma m}({ \bf R},t) \right) \label{eq:mot}
\ee
where, $q_{0}^{2} \equiv \langle V''(\sigma) \rangle = V''(\psi_{0}) + {\cal
O}(1/L)$ and $C_{\sigma m}({\bf R},t) = \langle \sigma (1) m(2) \rangle$.
Here,  equal times are considered and statistical homogeneity of the system has
been assumed so $ t_{1} = t_{2} = t $ and $ { \bf R = R_{2} - R_{1} }$. Since
\begin{eqnarray}
C({\bf R},t) & = & \langle \delta \psi(1) \delta \psi(2) \rangle \nonumber \\
& = & \langle \sigma(1) \sigma(2) \rangle - M^{2} + { \cal{O}} \left(
\frac{1}{L} \right)
\end{eqnarray}
we see that (\ref{eq:mot}) is an equation for $C({ \bf R},t) $ to  ${\cal O}
(1/L)$ since the action of the derivatives eliminates the disconnected part of
the correlation function.

A key aspect of the theory is the choice of the probability distribution $P[m]$
governing the field $m$. This point is discussed in some detail in references
\cite{maz,PGA1,physica article}. Here we limit ourselves to the case where
$P[m]$ is given by an off-set Gaussian with
\be
\bar{m}(t) = \langle m(1) \rangle_{0},
\ee
and $\langle \ldots \rangle_{0} $ is over a probability distribution
$P_{0}[\delta m]$ which is Gaussian with respect to $\delta m(1) = m(1) -
\bar{m}(t)$. $P_{0}[\delta m]$ is then determined by the variance
\be
C_{0} (12) = \langle \delta m(1) \delta m(2) \rangle_{0}.
\ee
Since the field $m$ can be approximately interpreted as the perpendicular
distance to the nearest interface, the off-set corresponds to a greater
probability to be in one phase than in the other. The effects of this non-zero
average will be explored in the next section.

%%%%%%%%%%%%%%%%%%%%%%%%%%%%%%%%%%%%%%%%%%%%%%%%%%%
\newcommand{\css}{\mbox{$C_{\sigma \sigma}$}}
\newcommand{\csm}{\mbox{$C_{\sigma m}$}}
\newcommand{\co}{\mbox{$C_{0} (12)$}}
\newcommand{\mb}{\mbox{$ \bar{m} (t)$}}
%%%%%%%%%%%%%%%%%%%%%%%%%%%%%%%%%%%%%%%%%%%%%%%%%%%
\section{Evaluation of Averages: the Scaling Equation of Motion }
\subsection{Evaluation of Averages}

In order for (\ref{eq:mot}) to be a closed equation for \css , it is necessary
to relate \csm\ to \css. As in \cite{maz}, this is done by using the Gaussian
nature of $m$. Now, however, the Gaussian probability distribution must satisfy
the condition $\langle m(1) \rangle_{0} = \mb \not= 0$. Taking this into
account one finds, using the standard properties of Gaussian integrals
\cite{TUG},
\begin{equation}
\csm (12) = \langle \sigma_{1} (1) \rangle_{0} \co + M \mb.   \label{eq:csm}
\end{equation}
The spatially independent term is eliminated by the action of the Laplacian in
(\ref{eq:mot}). Since $m$ is a Gaussian random field it follows that
\begin{equation}
\langle \sigma_{1} (1) \rangle_{0} = \int dx_1 \sigma_{1} ( x_{1} ) \Phi
(x_{1}) \label{eq:meansig1}
\end{equation}
where
\begin{equation}
\Phi (x_{1}) = \frac{1}{\sqrt{2 \pi S_{0} (t)}} e^{-\frac{1}{2} (x_{1} -
\bar{m}(t))^{2} / S_{0} (t) } \label{eq:phi}
\end{equation}
with
\begin{equation}
S_{0} (t) = C_{0} (11).
\end{equation}
Since $m$ is a measure of the distance away from an interface it is expected
that in the long-time scaling regime $S_{0} \sim L^{2}$ and $\mb \sim L $.
Therefore the limit
\begin{equation}
\lim_{t \rightarrow \infty} \frac{\mb}{\sqrt{S_{0} (t)}} = y
\end{equation}
exists. In evaluating (\ref{eq:meansig1}) it is important to note that, for a
wide class of potentials, $\sigma_{1}(x_{1})$ goes exponentially to zero for
large $|x_{1}|$. Therefore one can, after eliminating $\bar{m}$ in favour of $y
\sqrt{S_{0}(t)}$, expand (\ref{eq:meansig1}) in powers of $S_{0}^{-1}$ and use
(\ref{eq:intsig1}) to obtain
\begin{equation}
\langle \sigma_{1} (1) \rangle_{0}  = \frac{2 \psi_{0}}{\sqrt{2 \pi S_{0} (t)}}
e^{-\frac{1}{2} y^{2}} + {\cal O} ( S_{0}^{-1}).  \label{eq:s1}
\end{equation}
The parameter $y$ can be related to $M$ by using the derivative relation
\cite{mazenko-unpublished}
\begin{equation}
\frac{\partial M}{\partial \bar{m}(t)} = \frac{\partial \langle \sigma
\rangle_{0}}{\partial \mb} = \langle \sigma_{1} \rangle_{0},
\end{equation}
from which follows
\begin{equation}
M = \int^{\bar{m}(t)}_{0}dz \sqrt{\frac{2}{\pi S_{0}(t)}} \psi_{0} e^{-
\frac{1}{2} \frac{z^{2}}{S_0(t)}}.
\end{equation}
Thus, with $ \tilde{M} = M/ \psi_{0} $,
\begin{equation}
\tilde{M} = erf \left( y/ \sqrt{2} \right).  \label{eq:Mandy}
\end{equation}
Therefore, there is a one-to-one correspondence between $y$ and $\tilde{M}$.
$y=0$ is a critical
quench and $y \rightarrow \infty $ corresponds to a quench at the coexistence
curve.

With the definition
\begin{equation}
f({\bf R},t) = \frac{C_{0}( {\bf R} ,t)}{S_{0}(t)}
\end{equation}
and the use of (\ref{eq:csm}) and (\ref{eq:s1}), the equation of motion
(\ref{eq:mot}) takes the form
\begin{equation}
\frac{1}{2} \frac{\partial}{\partial t} \css( { \bf R},t) = - D_{0}q_{0}^{2}
\nabla^{2} \left( \frac{u_{0}}{L} \css ( { \bf R},t)+ \lambda
\sqrt{\frac{2S_{0}}{\pi}} \psi_{0} e^{- \frac{1}{2} y^{2}} \nabla^{2} f ( { \bf
R},t)\right). \label{eq:newmot}
\end{equation}
The theory can be closed by relating $C_{0} ( { \bf R},t) $ to $C_{\sigma
\sigma} ( { \bf R},t)$ via the relation \cite{TUG}
\begin{equation}
C_{11} ({ \bf R}, t) = \frac{\partial \css( { \bf R},t)}{\partial C_{0}( { \bf
R},t)} \label{eq:c11der}
\end{equation}
where $ C_{nm} ({ \bf R},t) = \langle \sigma_{n} (1) \sigma_{m} (2) \rangle_{0}
$. The fact that $m$ is Gaussian enables one to write
\begin{equation}
C_{11} ({ \bf R},t) = \int dx_{1}dx_{2} \sigma_{1}(x_{1}) \sigma_{1}(x_{2})
\Phi (x_{1},x_{2}) \label{eq:c11}
\end{equation}
where
\begin{equation}
\Phi (x_{1},x_{2}) = \frac{\gamma}{2 \pi S_{0}} e^{- \frac{1}{2}
\frac{\gamma^{2}}{S_{0}} [ (x_{1} - \bar{m}(t))^{2} + (x_{2} - \bar{m}(t))^{2}
-2f(x_{1} - \bar{m}(t))(x_{2} - \bar{m}(t))]}
\end{equation}
with $ \gamma = ( 1- f^{2})^{- \frac{1}{2}} $.
Again, $\bar{m}(t)$ can be eliminated in favour of $y \sqrt{S_{0}(t)}$ and the
result (\ref{eq:c11}) expanded in terms of $S_{0}^{-1}$. At long times the
leading order term dominates, resulting in
\begin{equation}
C_{11} ({ \bf R},t ) = \frac{\gamma}{2 \pi S_{0}} e^{\frac{-y^{2}}{1+f}} \int
dx_{1}dx_{2} \sigma_{1}(x_{1}) \sigma_{1}(x_{2}) +{ \cal O} (S_{0}^{-3/2})
\label{eq:formc11}
\end{equation}
where the integral is easily found to give $4 \psi_{0}^{2} $.
Integration of (\ref{eq:c11der}) with the substitution (\ref{eq:formc11}),
keeping in mind the definition of $f$,  gives the desired relation between
$C_{\sigma \sigma} \mbox{ and } f$:
\begin{equation}
\css = \frac{2 \psi_{0}^{2}}{\pi} \int_{0}^{f} \frac{ds}{\sqrt{1- s^{2}}}
e^{\frac{-y^{2}}{1+s}}.  \label{eq:csstof}
\end{equation}
For a critical quench $y = 0$, and this reduces to the now standard expression
$C_{\sigma \sigma} = \frac{2}{\pi} \psi_{0}^{2} \sin^{-1} f$.
Equations (\ref{eq:csstof}) and (\ref{eq:newmot}) form a closed system of
equations for \css\ .
\subsection{The Scaling Regime}

At long times the correlation function is expected to have a scaling solution.
The ansatz
\begin{equation}
\css ({\bf R},t) = \psi_{0}^{2} F(x),
\end{equation}
with $ x = |{\bf R}|/L(t)$ being the scaled distance, can be substituted into
equations (\ref{eq:newmot}) and (\ref{eq:csstof}) to give
\begin{eqnarray}
x \bar{F}' & = & \nabla^{2} ( \bar{u} \bar{F} + \nabla^{2} f) \label{eq:motfin}
\\
\bar{F} & = & \frac{2}{\pi} \int_{0}^{f} \frac{ds}{\sqrt{1 - s^{2}}} e^{ -
\frac{1}{2} y^{2} \frac{1-s}{1+s}} \label{eq:fonf}
\end{eqnarray}
where
\begin{eqnarray}
\bar{F} (x) & = & e^{ y^{2}/2} F(x) \\
\bar{u} & = & \frac{2 u_{0} D_{0} q_{0}^{2}}{L^{2} \dot{L}}
\end{eqnarray}
and we have chosen $L = ( \lambda \bar{u}/ \psi_{0} u_{0}) \sqrt{2 S_{0} / \pi}
$. Note that, for scaling we require that  $\bar{u}$ is a constant and
therefore $L \sim t^{1/3}$. One expects $\bar{F}$ to have a weaker $y$
dependence than $F$ for large $y$. It is useful to express $f$ in terms of
\begin{equation}
\Phi = \sqrt{ \frac{1-f}{1+f} } \label{eq:phiandf}
\end{equation}
in equation (\ref{eq:fonf}) to obtain the well-behaved integral representation
\begin{equation}
\bar{F} = \frac{4}{ \pi} \int_{ \Phi}^{1} \frac{ds}{1+s^{2}} e^{-
\frac{y^{2}s^{2}}{2}}. \label{eq:Fwithphi}
\end{equation}
For future reference, the spatial derivative of (\ref{eq:Fwithphi}) is
\begin{equation}
\frac{\partial \bar{F}}{\partial x} = - \frac{4}{ \pi} \frac{ e^{-y^{2}
\Phi^{2}/2}}{1 + \Phi^{2}} \frac{\partial \Phi}{\partial x} \label{eq:derfphi}.
\end{equation}
It is noteworthy that $\bar{F}$ has a lower bound. Let $\Phi \rightarrow
\infty$ in (\ref{eq:Fwithphi}) and write
\begin{equation}
\bar{F}_{min} = \frac{4}{ \pi} e^{y^{2}/2} \int_{ \infty}^{1}
\frac{ds}{1+s^{2}} e^{-\frac{y^{2}}{2} (1 + s^{2})} = \frac{4}{\pi}
e^{y^{2}/2}I(y).
\end{equation}
$I(y)$ has the following properties:
\begin{eqnarray}
I(0) & = & - \frac{\pi}{4} \nonumber \\
\frac{d}{dy}I(y) & = & \frac{\pi}{2} \left[ 1 - erf \left( \frac{y}{\sqrt{2}}
\right) \right] \frac{d}{dy}  erf( \frac{y}{\sqrt{2}}).
\end{eqnarray}
Integrating these equations yields
\be
I(y) =  - \frac{\pi}{4} \left( 1 - erf \left( \frac{y}{\sqrt{2}} \right)
\right)^{2} = - \frac{\pi}{4}( 1 - |\tilde{M}|)^{2}.
\ee
Thus, the lower bound on $\bar{F}$ is
\be
\bar{F}_{min} = -e^{y^{2}/2}(1 - |\tilde{M}|)^{2}. \label{eq:Fmin}
\ee
As $\tilde{M} \rightarrow 1$ this lower bound approaches zero.

Equations (\ref{eq:motfin}) and (\ref{eq:Fwithphi}) constitute a non-linear
eigenvalue problem for $ \bar{F}$, with a unique solution determined by the
boundary conditions at small $x$ and the physical condition $\bar{F}
\rightarrow 0 \mbox{ exponentially as } x \rightarrow \infty$. We will see that
$\bar{u}$ is determined as part of the solution. The only parameters entering
into the determination of $ \bar{F} $ are $y$ and the dimension d, which
appears in the spherically symmetric
Laplacian.
\section{Limiting Cases}
\subsection{Small x behaviour}

The small $x$ behaviour of $\bar{F}$ can be determined analytically. We find
that $\bar{F} $  has the form
\begin{equation}
\bar{F} = \bar{F}(0) - \alpha x (1 + \beta_{2} x^{2} + \beta_{3} x^{3} + \cdots
).
\end{equation}
Expanding $ \Phi$ for small $x$
\begin{equation}
\Phi = \Phi_{1} x + \Phi_{2} x^{2} + \cdots
\end{equation}
and using (\ref{eq:derfphi}) to connect the power series for $\bar{F}$ and
$\Phi$ gives
\begin{eqnarray}
\Phi_{1} & =  & \frac{ \pi \alpha}{4} \label{eq:Phi1} \\
\Phi_{2} & = & \frac{ \pi \alpha \beta_{2}}{4}. \label{eq:Phi2}
\end{eqnarray}
Substitution of these results into (\ref{eq:motfin}) give, at ${\cal O} (1/x)$
\be
\bar{u} = \frac{-3 \pi^{2} \alpha \beta_{2} (d+1)}{4}.
\ee
$ \bar{F} (0) $ can be determined from (\ref{eq:Fwithphi}) by noting that $
\Phi (0) = 0 $ and then using the same technique that was used to derive
(\ref{eq:Fmin}) . The result is
\be
\bar{F} (0) = (1- \tilde{M}^{2}) e^{y^{2}/2}.
\ee
The equation of motion (\ref{eq:motfin}) can be partially integrated using the
Green's function for the Laplacian \cite{mazenko-unpublished}. For $ d > 2 $
the result is
\be
\bar{u} \bar{F} + \nabla^{2} f = \frac{1}{d-2} \left[ \frac{d}{x^{d-2}}
I_{d}(x) - 2( I_{2}(x) - I_{2}( \infty)) \right] \label{eq:intmot}
\ee
with
\begin{equation}
I_{d} (x)   =  \int_{0}^{x} z^{d-1} \bar{F}(z) dz.  \label{eq:I}
\end{equation}
The results (\ref{eq:Phi1}) and (\ref{eq:Phi2}) can be substituted into
(\ref{eq:intmot}). Since one can show that
\be
\lim_{x \rightarrow 0} \frac{ I_{d}(x)}{x^{d-2}}  = \lim_{x \rightarrow 0}
I_{2}(x) = 0
\ee
one has at ${\cal O}(1)$
\be
\frac{2}{d-2} I_{2} ( \infty) = - \frac{ \pi^{2} \alpha^{2} d}{4} + \bar{u}
\bar{F} (0).
\ee
For $d=2$ the analogous results are
\be
\bar{u} \bar{F} + \nabla^{2} f = (1 - 2 \ln{x})I_{2}(x) + 2(J_{2}(x) -
J_{2}(\infty)) \label{eq:d=2}
\ee
with
\be
J_{2}(x) = \int^{x}_{0} dz \mbox{ } z \bar{F}(z) \ln{z}.
\ee
Since $\lim_{x \rightarrow 0} J_{2}(x) =0$ substitution of (\ref{eq:Phi1}) and
(\ref{eq:Phi2}) into (\ref{eq:d=2}) give, at ${\cal O}(1)$
\be
-2J_{2}(\infty) = - \frac{\pi^{2} \alpha^{2} }{2} + \bar{u} \bar{F}.
\ee
Knowledge of the parameters $\alpha$ and $\beta_{2}$ allows one to determine
the constants $\bar{u}$ and $I_{2}( \infty)$ (or $J_{2}(\infty)$) appearing in
the equation of motion for a given value of $y$. Numerically, what this means
is that values for $\alpha$ and $\beta_{2}$ are chosen and equations
(\ref{eq:intmot}) (or (\ref{eq:d=2})) and (\ref{eq:Fwithphi}) integrated
forward in $x$. $\alpha$ and $\beta_{2}$ are adjusted until $\bar{F}$ satisfies
both the conservation law (\ref{eq:conslaw}) and the boundary condition
$\bar{F} \rightarrow 0 \mbox{ exponentially as } x \rightarrow \infty $.
\subsection{Large x behaviour}

For large $x$ both $\bar{F}$ and $f$ are small and $\Phi$ can be expanded about
its asymptotic value
\be
\Phi (x) = 1 + \eta (x).
\ee
Substitution into (\ref{eq:derfphi}) and integration yields a relation between
$ \bar{F}$  and $ \eta$
\be
\bar{F}(x) = - \frac{2}{ \pi} e^{-y^{2}/2} \eta (x).
\ee
Also, from (\ref{eq:phiandf}) we have $ f(x) = - \eta (x)$ so we may rewrite
(\ref{eq:motfin}) in the large-$x$ limit as
\be
x \bar{F}' = \nabla^{2}(\bar{u} \bar{F} + \frac{\pi}{2} e^{y^{2}/2} \nabla^{2}
\bar{F}).
\ee
This equation supports damped oscillatory solutions of the form
\be
\bar{F} \sim  \bar{F}_{0} x^{-2d/3} \exp \left( - \frac{3 e^{-
y^{2}/6}}{2^{8/3} \pi^{1/3}} x^{4/3} \right) \cos \left( \frac{3^{3/2} e^{-
y^{2}/6}}{2^{8/3} \pi^{1/3} } x^{4/3} + \phi \right)
\ee
where $ \bar{F}_{0} \mbox{ and } \phi $ are constants. Note that as $y
\rightarrow \infty $ the  wavelength of the oscillations increases and the
exponential term goes to 1. This means that in this limit one must go to
progressively larger values of $x$ before one sees this asymptotic behaviour.
\subsection{Small Q  behaviour}

It is the Fourier transform of the order parameter correlation function,
\begin{eqnarray}
C({\bf q},t) & = & \int d^{d}R e^{i {\bf q \cdot R}} \langle \psi({\bf R},t)
\psi({\bf 0},t) \rangle \nonumber \\
             & = & \psi_{0}^{2} L^{d}(t) \tilde{F}(Q) + (2\pi)^{d}M^{2}
\delta(q)
\end{eqnarray}
with $Q = qL(t)$ and it is
\be
\tilde{F}(Q) = \int d^{d}x e^{i {\bf Q \cdot x}} F(x),
\ee
that is measured in a scattering experiment. In the total scattering
cross-section we expect that at long times there is a dynamic contribution to
the forward Bragg peak,
\be
\lim_{t \rightarrow \infty} \psi_{0}^{2} L^{d}(t) \tilde{F}(Q) = A \delta(q),
\ee
in addition to the static contribution $(2\pi)^{d} M^{2} \delta(q)$. Since
\begin{eqnarray}
A & = & \int d^{d}q \mbox{ } \lim_{t \rightarrow \infty} \psi_{0}^{2} L^{d}(t)
\tilde{F}(Q) \nonumber \\
& = & (2\pi)^{d} \psi_{0}^{2} F(0) = (2\pi)^{d} \psi_{0}^{2} (1 -
\tilde{M}^{2}),
\end{eqnarray}
the total contribution to the forward Bragg peak at late times is $(2\pi)^{d}
\psi_{0}^{2} \delta(q)$, as expected.

To examine the small $Q$ behaviour of the structure factor $\tilde{F}(Q)$, it
is useful to consider the moments of $F(x)$
\be
W_{p} = \int d^{d}x x^{p} F(x)
\ee
which can be found by multiplying (\ref{eq:motfin}) by $x^{p}$ and integrating.
The result is that $ W_{0} = W_{2} = 0 $ while
\be
W_{4} = - \frac{8d(d+2)}{d+4} e^{- \frac{1}{2} y^{2}} \int d^{d} x f(x).
\ee
Thus we have
\be
\tilde{F}(Q) \sim AQ^{4}
\ee
for small $Q$ where
\be
A = - \frac{e^{-y^{2}/2}}{d + 4} \int d^{d} x f(x). \label{eq:A}
\ee
The $Q^{4}$ behaviour of $ \tilde{F} (Q) $ at small $Q$ is a consequence of the
fact that in the theory $u$ is conserved away from the interfaces, and this
behaviour does not depend on the specific ansatz for u. In order for $
\tilde{F} (Q) $ to be positive definite (\ref{eq:A}) implies that
\be
\int d^{d} x f(x) < 0.
\ee
This means that $\lim_{q \rightarrow 0} \langle |m_{q}(t)|^{2} \rangle / S_{0}
< 0 $. As pointed out by Yeung {\em et al}. \cite{Yeung} this is a shortcoming
of the fact that $m$ is a Gaussian variable. This problem is resolved when
non-Gaussian corrections to the probabilty distribution for the field $m$ are
considered \cite{maz}.
\subsection{Large y behaviour}

An analytic result for the limit as one approaches the coexistence curve,
$\tilde{M} \rightarrow 1 \mbox{ , } y \rightarrow \infty$ is of interest
because it allows one to make comparisions with other theories developed for
this regime. From the numerical analysis in the next section the following
facts emerge. The first is that, as $y$ increases, the scaled length $x$ over
which the correlation function takes significant values decreases. This
suggests that $x$ should be rescaled as
\be
x = y^{p}z \label{eq:scalx}
\ee
with $p < 0$. Second, it appears that $\bar{u}$ grows as some power of $y$ for
large $y$. We are led to assume the form
\be
\bar{u} = \bar{u}_{\infty} y^{n} \label{eq:upower}
\ee
with $ n > 1$. Finally, since the interesting behaviour occurs near the origin
where the quantity $\Phi$ is small, we can
rescale $\Phi$
\be
\Phi = \frac{\phi}{y} \ll 1 \label{eq:scalphi}
\ee
where $\phi$ grows slowly but does not break the bound near the origin. Using
this definition in equation (\ref{eq:phiandf}) leads to
\be
f \approx 1 - 2 \frac{\phi^{2}}{y^{2}} \mbox{ }.
\ee

Armed with these results we proceed to re-examine the theory. Using
(\ref{eq:scalphi}) in  equation (\ref{eq:Fwithphi}) and letting $t = y s$ in
the integrand gives
\be
\bar{F}(z) = \frac{4}{\pi y} \int_{\phi(z)}^{y} \frac{dt}{1+(t/y)^{2}}
e^{-t^{2}/2}.
\ee
To leading order in $y^{-1}$ this is
\begin{equation}
\bar{F}(z)  =  \frac{4}{\pi y} \int_{\phi(z)}^{\infty} dt e^{-t^{2}/2}
            =  \frac{1}{y} \bar{F}_{\infty} (z) \label{eq:Fasym}
\end{equation}
where
\be
\bar{F}_{\infty} (z) = 2 \sqrt{\frac{2}{\pi}} \left( 1 - erf \left( \frac{\phi
(z)}{\sqrt{2}} \right) \right). \label{eq:finfty}
\ee
Under the rescaling outlined above the equation of motion (\ref{eq:intmot}) for
$d>2$ becomes
\be
\bar{u}_{\infty} y^{n-1} \bar{F}_{\infty} (z) - 2 y^{-2(p+1)} \nabla_{z}^{2}
\phi^{2} = \frac{y^{2p-1}}{d-2} \left[ \frac{d}{z^{d-2}}I_{d}(z) - 2( I_{2}(z)
- I_{2}( \infty) ) \right] \label{eq:mess}
\ee
with
\be
I_{d}(z) = \int_{0}^{z} ds s^{d-1} \bar{F}_{\infty} (s).
\ee
Since $p < 0$ the right hand side does not contribute as $y \rightarrow
\infty$. The integrals in the $d=2$ case also do not contribute in this limit
so the following results are valid for $d \geq 2$. Balancing powers of $y$ on
the left hand side of (\ref{eq:mess}) gives
\be
p = - \frac{n+1}{2} \label{eq:pandn}
\ee
and
\be
\bar{u}_{\infty} \bar{F}_{\infty} - 2 \nabla^{2} \phi^{2} = 0.
\ee
Using (\ref{eq:finfty}) this becomes a simple equation for $\phi$
\be
\bar{u}_{\infty} \sqrt{\frac{2}{\pi}} \left( 1 - erf \left(
\frac{\phi}{\sqrt{2}} \right) \right) = \nabla^{2} \phi^{2}.
\label{eq:largeylim}
\ee
In the numerical solution of (\ref{eq:largeylim}), $ \bar{u}_{ \infty} $ is a
parameter which is found from a fit of the numerically determined $\bar{u}$ to
the form (\ref{eq:upower}) for large $y$.
\subsection{Small and large $z$ behaviour for large $y$}

An examination of (\ref{eq:largeylim}) in the limit of large and small $z$ is
instructive. When $z$ is small $\phi$ is expected to be small so $\phi$ can be
expanded as a power series in $z$
\be
\phi = \phi_{1} z + \phi_{2} z^{2} + \cdots.
\ee
Also, in this limit, (\ref{eq:largeylim}) simplifies to the form
\be
\bar{u}_{\infty} \sqrt{\frac{2}{\pi}} \left( 1 - \sqrt{\frac{2}{\pi}} \phi
\right) = \nabla^{2} \phi^{2}.
\ee
Matching powers of $z$ to leading order gives
\be
\phi_{1} = \left( \frac{2}{\pi} \right)^{1/4} \left(
\frac{\bar{u}_{\infty}}{2d} \right)^{1/2} \mbox{ .}
\ee
This result allows one to make predictions about the asymptotic behaviour of
$\alpha$ and $\beta_{2}$ when $y$ is large. Using (\ref{eq:scalx}) and
(\ref{eq:finfty}) one can write
\be
\bar{F}(x) = \frac{2}{y} \sqrt{\frac{2}{\pi}} - \frac{4}{\pi} \phi_{1} y^{-p-1}
x + \cdots
\ee
giving
\be
\alpha = \alpha_{\infty} y^{-p-1}
\ee
for large $y$ with $\alpha_{\infty} = 4 \phi_{1} / \pi$. If we also write
\be
\beta_{2} = \beta_{2 \infty} y^{m}
\ee
then
\be
\bar{u} = - \frac{3 \pi^{2} \alpha_{ \infty} \beta_{2 \infty} (d +1)}{4}
y^{m-p-1} = \bar{u}_{\infty} y^{n}.
\ee
Matching the coefficient and the exponent leads, using (\ref{eq:pandn}), to the
following relationships:
\begin{eqnarray}
\alpha & = & \alpha_{\infty} y^{(n-1)/2} \label{eq:algy}\\
\beta_{2} & = & \beta_{2 \infty} y^{(n+1)/2} \label{eq:blgy}\\
\alpha_{\infty} & = & - \frac{12}{\sqrt{2 \pi}} \left( \frac{d+1}{d} \right)
\beta_{2 \infty}. \label{eq:ratioatob}
\end{eqnarray}
Thus, a graph of $\beta_{2}/ \alpha$ vs. $y$ for large $y$ will be linear with
a slope that depends only on the dimensionality of the system.

At large $z$, $\phi$ is large and (\ref{eq:largeylim}) is well approximated by
\be
\bar{u}_{\infty} \frac{2}{ \pi \phi} e^{- \phi^{2}/2} = \nabla^{2} \phi^{2}.
\ee
For $d > 2$, standard asymptotic analysis yields, at next to leading order:
\be
\phi (z) = \ln^{1/2} \left( \frac{z^{4}}{\ln{z}} \left(
\frac{\bar{u}_{\infty}}{2 \pi (d-2)} \right)^{2} \right).
\ee
This implies that
\be
\bar{F}_{ \infty} (z) = \frac{4(d-2)}{\bar{u}_{\infty}} \frac{1}{z^{2}}
\ee
for large $z$ and $d > 2$.
For $d=2$ one has
\be
\bar{F}_{ \infty} (z) = \frac{ \bar{F}_{0}}{(1+ \frac{\bar{u}_{\infty}
\bar{F}_{0}}{16} z^{2})^{2}}.
\ee
It is clear that, near the coexistence curve, the oscillations in the scaling
form become insignificant and are dominated by a strong decay.
%%%%%%%%%%%%%%%%%%%%%%%%%%%%%%%%%%%%%%%%%%%%%%%%%%%%%%%%%%%%%%%%%%%%%%%%%%%%%%

\section{Numerical Solution of the Non-linear Eigenvalue Problem}
\subsection{$\alpha$ and $\beta_{2} $ as a function of $y$ }

In this section the numerical solution of (\ref{eq:intmot}) (or (\ref{eq:d=2}))
coupled with (\ref{eq:Fwithphi}) in two and three dimensions will be discussed.
The equations are integrated forward from $x=0$ using a fourth order
Runge-Kutta integrator with step size $ \delta x = 0.001$, subject to the
initial conditions $ \bar{F}(0) = (1- \tilde{M}^{2})e^{y^{2}/2}$ , $
\bar{F}'(0) = - \alpha $ , $ \bar{F}''(0) = -2 \alpha \beta_{2} $, and
$f'(0)=0$. This method of integration seems numerically stable and insensitive
to the choice of $ \delta x$. The search for the eigenvalues $ \alpha$ and $
\beta_{2}$ involves requiring that the solution $ \bar{F}$ obey the
conservation law (\ref{eq:conslaw}) and have the physically acceptable
behaviour $ \bar{F} \rightarrow 0 \mbox{ exponentially as } x \rightarrow
\infty $. This search is performed by fixing $ \alpha$ and then searching for
the value of $ \beta_{2} $ which pushes the diverging, unphysical solution to
larger values of $x$. The value of $ \alpha$ is then adjusted so that the flat
region of $ \bar{F}$ at large $x$ is properly zeroed. The procedure is repeated
with the new value of $ \alpha$ until the exponentially growing solution is
pushed as far from the origin as possible and until $ \bar{F}$ is zeroed as
well as possible. The degree to which the conservation law is satisfied
naturally depends on how well the function is zeroed. The convergence of the
eigenvalues is fast and $ \bar{F}$ can be zeroed to better than $10^{-6}$ using
this method.

The results for the eigenvalues $ \alpha$ and $ \beta_{2}$  are shown in Fig.
\ref{fig:alpha}. One sees that $ \alpha$ initially decreases reaching a minimum
at $y \sim 1$ and then rapidly becomes large and positive as the coexistence
curve is approached. The eigenvalue $\beta_{2}$ is negative at $y = 0$ and
monotonically decreases as $y $ increases, decreasing rapidly as $y$ becomes
large $( \tilde{M} \rightarrow 1)$.  Equation (\ref{eq:ratioatob}) predicts
that a graph of $\beta_{2}/\alpha$ will be linear for large $y$ with a slope
-0.157 for $d=3$ and -0.139 for $d=2$ . We see this linear behaviour and have
measured slopes of about -0.143 and -0.129 for $d=2 \mbox{ and } 3$
respectively. The exponents $n \mbox{ and } p$, and the coefficient
$\bar{u}_{\infty}$ defined in (\ref{eq:scalx}) and (\ref{eq:upower}) can be
found by fitting the large $y$ behaviour of $\alpha$ and $\beta_{2}$ to the
forms (\ref{eq:algy}) and (\ref{eq:blgy}). For three dimensions one finds $n
\sim 8 \mbox{, } p \sim -4.5 \mbox{ and } \bar{u}_{\infty} \sim 0.0033$. In two
dimensions one has $n \sim 6 \mbox{, } p \sim -3.5 \mbox{ and }
\bar{u}_{\infty} \sim 0.024$. When considering these results it should be kept
in mind  that only a few values of $y$ around $y=4$  were used to obtain these
values. In principle, both the exponents and the coefficient can be accurately
obtained by extending the numerical analysis to larger values of $y$. In
practice, this is difficult due to reasons that are discussed below.
\subsection{Scaling Function as a function of $ \tilde{M} $}

The dependence of the scaling function $F(x)$ on $ \tilde{M} $ is shown in Fig.
\ref{fig:FsmM2} for two dimensions and in Fig. \ref{fig:FsmM} for three
dimensions. In these plots $F(x)$ is normalized so that $F(0) = 1$. Both
figures show  that $F(x)$ depends only weakly on $ \tilde{M} $ for values of $
\tilde{M} < 0.4$. The scaling function has a prominent oscillatory component
which is necessary to satisfy the conservation law. At intermediate values of $
\tilde{M}$, the position of the first minimum of $F(x)$ occurs at larger values
of $x$ and the depth of this minimum decreases as $\tilde{M}$ increases. The
depth of the oscillations is greater in two dimensions than in three. These
stronger oscillations make the presence of the lower bound on $F(x)$
noticeable, and near the coexistence curve the minima in the scaling function
are very flat in order to be consistent with this bound.

As $\tilde{M} \rightarrow 1$ the scaling function approaches its asymptotic
form (\ref{eq:finfty}) which can be determined by numerically solving
(\ref{eq:largeylim}) using the values of $\bar{u}_{\infty}$ found in the
previous section. Since we know the exponent $p$, we can rescale the distance
$x$ using (\ref{eq:scalx}) and plot $\bar{F}_{\infty}(z) = y \bar{F}(xy^{-p})$
for large values of $y$. This is done for two and three dimensions in Fig.
\ref{fig:FlargeM}. The asymptotic forms obtained by solving
(\ref{eq:largeylim}) are also shown in this figure. We see that $F(x)$ decays
very rapidly when the system is near the coexistence curve. Oscillations do
occur for these values of $y$, but they occur at large  $x$ and have a small
amplitude and large wavelength. For $d =3$ the curves appear to approach the
asymptotic form as $y$ increases. Hovever, for $d=2$ the asymptotic form is not
approached if one uses  $\bar{u}_{\infty} = 0.024$. A value of
$\bar{u}_{\infty} = 0.036$ gives a better fit and the form obtained using this
value is the one shown in Fig. \ref{fig:FlargeM}. Matching the asymptotic form
to the rescaled large $y$ scaling function is another way to determine
$\bar{u}_{\infty}$. We believe that the two methods for finding  $
\bar{u}_{\infty}$ give different values because in the fit of  $\alpha$ and
$\beta_{2}$ to the forms (\ref{eq:algy}) and (\ref{eq:blgy}) we do not have
values of $y$ which are large enough to be in the asymptotic regime. Larger
values of $y$ are difficult to reach because one runs into numerical problems
as the theoretical lower bound on $F(x)$ approaches zero. These numerical
problems are  especially significant in two dimensions since the oscillations
in the correlation function are stronger than in three dimensions.
\subsection{Scaling of the Structure Factor}

The structure factor,
\be
\tilde{F}(Q) = \int d^{d}x e^{i {\bf Q \cdot x}} F(x)
\ee
was calculated by taking the Fourier transform of our numerically determined
$F(x)$. We find that as $\tilde{M}$ increases the height of the peak decreases
and the peak position moves to smaller values of $Q$. Graphs of the normalized
structure factor for various $\tilde{M}$ are shown in Fig. \ref{fig:nsf2} for
$d=2$ and Fig. \ref{fig:nsf} for $d=3$. Logarithmic  plots reveal the power-law
dependence of $\tilde{F}(Q)$ for large and small $Q$ (Fig. \ref{fig:loglog}).
For small $Q$, $\tilde{F}(Q) \sim Q^{4}$ in both $d=2 \mbox{ and } 3$, for all
$\tilde{M}$.  Small deviations from the $Q^{4}$ behaviour can be seen, but we
atttribute these to the unreliablity of the numerical determination of $F(x)$
for extremely large $x$. For large $Q$, one observes Porod's law, $\tilde{F}(Q)
\sim Q^{-(d+1)}$, for all $\tilde{M}$. The coefficients $A_{4} \mbox{ and }
A_{P}$ defined in the introduction are determined and plotted in Figs.
\ref{fig:A4} and \ref{fig:AP} respectively as functions of $\tilde{M}$.
$A_{4}$  increases with increasing $\tilde{M}$ and   $A_{P}$ is a decreasing
function of $\tilde{M}$, approaching zero like  a cusp at the coexistence
curve.

 Figures \ref{fig:nsf2} and \ref{fig:nsf} show that the width of the peak
increases slightly, but is rather insensitive to changes in $\tilde{M}$ until
very near the coexistence curve. In the logarithmic plots there appear to be
damped oscillations in $\tilde{F}(Q)$ at intermediate $Q$ before the onset of
the $Q^{-(d+1)}$ behaviour. In two dimensions, as one approaches the
coexistence curve, the main peak decreases in amplitude until it is comparable
to these oscillations, which show up as a shoulder to the main peak.  In three
dimensions there is  a tail on the large $Q$ side of the peak in the structure
factor, which also grows as $\tilde{M}$ increases. Both the secondary peak and
the tail may be related to the fact that the Tomita sum rule is stongly broken
as $ \tilde{M} \rightarrow 1$. The large coefficent of $x^{2}$ in the small $x$
expansion of $ \bar{F} (x) $ will lead to corrections to Porod's law for the
medium $Q$ behaviour of the structure factor.
\section{Comparisons}

In order to test the validity of the assumptions made in this paper the results
for $F(x)$ and $\tilde{F}(Q)$ will be compared with the relevant results of
other investigators. Experiments involving neutron scattering off of a binary
alloy have been done for a fixed $\tilde{M}$  \cite{expt}, but we have been
unable to find any experimental study of the dependence of  $\tilde{F}(Q)$ on
$\tilde{M}$. One problem with doing experiments near the coexistence curve is
that the small volume fraction of the minority phase causes the structure
factor to have a small amplitude, thus making it difficult to measure. Another
problem that arises when comparing experiment to theory is that it is  unclear
what volume fraction was used in a given  experiment, making a straightforward
comparision difficult.

While there are high quality numerical simulations for critical quenches
\cite{Oono,Rogers,Gunton-sim}, there has been far less work on off-critical
quenches. One example is the direct numerical simulation of the Cahn-Hilliard
equation in two dimensions performed by Chakrabarti {\em et al.} \cite{chak}.
Here, we compare their result for the correlation function with ours. Their
functions are scaled so that the first zero of the correlation function occurs
at $x=1$, and we have adjusted our length scale to correspond to this. The
comparisions for volume fractions $\phi = 0.5, 0.21 \mbox{ and } 0.05$ are
shown in Fig. \ref{fig:chak}. The relationship between $ \tilde{M}$ and the
volume fraction $ \phi$ is
\be
\phi = \frac{1}{2} ( 1 - \tilde{M} )
\ee
which is valid for quenches to $ T = 0 $. The quantitative agreement is poor.
In particular, the theory predicts that at large $x$ the oscillations in $F(x)$
have much larger amplitude than seen in the simulations. Nevertheless, the
positions of peaks and troughs of the oscillations are in qualitative
agreement. In addition, we agree on the observation that oscillations in the
correlation function become weaker and have longer wavelength as the
coexistence curve is approached. In summary, the qualitative agreement is
reasonable. We are unaware of any direct simulation of the Cahn-Hilliard
equation in three dimensions for the off-critical case. Such simulations are
difficult because large system sizes are required to give a statistically
meaningful distribution of droplets when the volume fraction is small.

One can also  make comparisons with generalizations of the LSW theory
\cite{Rikvold,Marqusee,Tomita-model,Ohta,Tokuyama,Furukawa,Marder,Yao,Nakahara}
{}. This is in the regime of Ostwald ripening \cite{Ost}. While much of the
analysis in this case has focussed on the droplet distribution function, more
recently a number of authors have determined $\tilde{F}(Q)$. In particular,
here we will compare our three dimensional results with those of Akaiwa and
Voorhees \cite{Voorhees}. They assume that the droplets are spherical and
interacting essentially electrostatically through a concentration field with
both monopole and dipole contributions.  Both the droplet size distribution
function and the structure factor can be extracted by numerical simulation of
the equations produced by the theory. Our structure factors and those of
\cite{Voorhees} are compared in Fig. \ref{fig:Voorhees}.  Both results agree
and give Porod's law at large $Q$. At small $Q$ both results exhibit $Q^{4}$
behaviour, although our results seem to have a smaller coefficient of $Q^{4}$
 than theirs. This may also be why the widths of our peaks are consistently
smaller than those of \cite{Voorhees}. There is also significant disagreement
on the shape of the structure factor for values of $Q$ just above the peak. In
the theory presented here this regime of $Q$ may be strongly affected by the
breaking of the Tomita sum rule.

\section{Conclusion}

In this paper it has been shown that the theory developed in \cite{maz} can be
extended to the case of off-critical quenches. The LSW $t^{1/3} $ law and the
associated scaling behaviour are determined for the entire concentration range.
The scaling function is a function only of the parameters $d$ and $ \tilde{M}
$, changing significantly only close to the coexistence curve where the
oscillations observed in the critical case are damped out. The structure factor
exhibits Porod's law for large $Q$ and $Q^{4}$ behaviour at small $Q$. This is
the first theory which is capable of sensibly treating spinodal decomposition
over the entire concentration range.

As discussed above, there are a number of virtues of this theory. However,
there are also important limitations. First, we have not been able to make
contact with the LSW theory in the $\tilde{M} \rightarrow 1$ limit. This will
require extending the current theory (or some improved version) to treat the
droplet
distribution in the dilute limit. This is a difficult but, to us, interesting
challenge. Secondly, it is clear that we must extend the theory developed here
to include non-Gaussian corrections if we are to remedy the problem of
$C_{0}(q,t)$ going negative for small $q$. Since one expects this quantity to
enter the determination of the droplet distribution function in an important
way it is crucial to include non-Gaussian corrections if one is to make
progress in this area. Non-Gaussian corrections have already been used to treat
the critical COP case and this is  discussed in \cite{maz}. Finally, it seems
reasonable to assume that the primary reason that we do
not obtain good quantitative agreement for $\tilde{F} (Q)$ and $F(x)$ is that
we do not satisfy the Tomita sum rule. We speculate that in order to satisfy
the Tomita sum rule an improved treatment of the gradient term in the
consituitive relation (\ref{eq:const}) is required.

\acknowledgements

This work was supported by the NSF through Grant No. NSF-DMR-91-20719. The
authors would like to thank Amitabha Chakrabarti, James Gunton, Norio Akaiwa,
and Peter Voorhees for providing them with the data used in the comparisions.
One of the authors (R. W.) acknowledges support from NSERC Canada.

\begin{figure}
\caption{ The eigenvalues $\alpha$ and  $\beta_{2}$ as a function of $y$. (a)
$\alpha$ for two dimensions (upper curve) and threedimensions (lower curve).
(b) $\beta_{2}$ for two dimensions (lower curve at $y=0$) and three dimensions
(upper curve at $y=0$). \label{fig:alpha}}
\end{figure}

\begin{figure}
\caption{ The normalized scaling function $F(x)$ in two dimensions for various
$\tilde{M}$. In terms of decreasing depth of the first minimum the curves
correspond to (a) $\tilde{M} =$ 0, 0.2, and 0.4. (b) $\tilde{M} =$ 0.6, 0.8,
and 0.9.  \label{fig:FsmM2}}
\end{figure}

\begin{figure}
\caption{ The normalized scaling function $F(x)$ in three dimensions for
various $\tilde{M}$. In terms of decreasing depth of the first minimum the
curves correspond to (a) $\tilde{M} =$ 0, 0.2, and 0.4. (b) $\tilde{M} =$ 0.6,
0.8, and 0.9.  \label{fig:FsmM}}
\end{figure}

\begin{figure}
\caption{ The large $y$ asymptotic scaling function $\bar{F}_{\infty}(z)$. From
lowest to uppermost the solid curves correspond to $y$ = 2.5, 3, 3.5, and 4
($\tilde{M}$ = 0.9876, 0.9973, .9995, and  .9999 respectively). The infinite
$y$ form ($\tilde{M} = 1$), obtained from our asymptotic analysis is shown as a
dashed line. (a) two dimensions. (b) three dimensions.  \label{fig:FlargeM}}
\end{figure}

\begin{figure}
\caption{The normalized structure factor in two dimensions. From lowest to
uppermost the curves correspond to  $\tilde{M}$ = 0, 0.4, 0.6, 0.8, and 0.9.
\label{fig:nsf2}}
\end{figure}

\begin{figure}
\caption{The normalized structure factor in three dimensions. From lowest to
uppermost the curves correspond to $\tilde{M}$ = 0, 0.4, 0.6, 0.8, and 0.9.
\label{fig:nsf}}
\end{figure}

\begin{figure}
\caption{Logarithmic plots of $e^{y^{2}/2}\tilde{F}(Q)$. At $\ln(Q)=0$, from
lowest  to uppermost the curves correspond to $\tilde{M} =$ 0, 0.4, 0.6, 0.8,
and 0.9.  (a) two dimensions. (b) three dimensions. In both graphs $Q^{4}$
behaviour is seen at small $Q$  and Porod's law $Q^{-(d + 1)}$ behaviour occurs
at large $Q$ (dotted lines). \label{fig:loglog}}
\end{figure}

\begin{figure}
\caption{The coefficient $A_{4} (\tilde{M})$ appearing in the small $Q$
behaviour of $\tilde{F} (Q)$ for two (upper curve) and three (lower curve)
dimensions. \label{fig:A4}}
\end{figure}

\begin{figure}
\caption{The Porod's law coefficient $A_{P} (\tilde{M})$ for two (lower curve)
and three (upper curve) dimensions. \label{fig:AP}}
\end{figure}

\begin{figure}
\caption{A comparison of our scaling forms for the correlation function (solid
lines) in two dimensions with those of \protect\cite{chak} (dashed lines). The
horizontal axis has been chosen so that the first zero of $F(x)$ occurs at
$x=1$ for both functions. (a) $\phi = 0.5$. (b) $\phi = 0.21$. (c) $\phi =
0.05$. \label{fig:chak}}
\end{figure}

\begin{figure}
\caption{A comparison  of our structure factors (solid lines)  in three
dimensions with those of \protect\cite{Voorhees} (dashed lines). (a) $\phi =
0.3$. (b) $\phi = 0.2$. (c) $\phi = 0.1$. (d) $\phi  = 0.05$.
\label{fig:Voorhees}}
\end{figure}


\begin{thebibliography}{99}
\bibitem{asym pot} Our results are only dependent on the fact that $V(\psi)
\rightarrow \infty \mbox{ as } |\psi| \rightarrow \infty$ and the fact that
there are  two degenerate minima. For simplicity, we have chosen to examine a
system with a symmetric potential, but the results of this paper are unchanged
if one considers an asymmetric potential. If the potential has two degenerate
minima at $\psi_{+} \mbox{ and } \psi_{-}$ with $ \psi_{+} > \psi_{-}$ one has
only to redefine $\psi_{0} = \frac{1}{2}( \psi_{+} - \psi_{-})$ and $\tilde{M}
= (M - \frac{1}{2}(\psi_{+} + \psi_{-}))/\psi_{0}$ to use the results of this
paper.
Note that in the symmetric case $\psi_{+} = - \psi_{-}$ and these definitions
coincide with those given in the body of this paper.
\end{thebibliography}
\end{document}